\documentclass[aps,prc,groupedaddress,showpacs,manuscript]{revtex4-1}

\usepackage{amsmath}
\usepackage{graphicx}
\usepackage{colordvi}
\usepackage{ulem}

\begin{document}

\title{Non-equilibrium dynamics in heavy ion collisions at low SIS energies}

\author {Qingfeng Li$\, ^{1}$\footnote{E-mail address:
liqf@hutc.zj.cn}, Caiwan Shen$\, ^{1}$, Chenchen Guo$\, ^{1,2}$,
Yongjia Wang$\, ^{1,3}$, Zhuxia Li$\, ^{4}$, J. Lukasik$\, ^{5,6}$,
and W. Trautmann$\, ^{5}$} \affiliation{ 1) School of Science,
Huzhou Teachers College, Huzhou 313000,
P.R. China \\
2) College of Physics Science $\&$ Technology, Shenyang Normal University, Shenyang 110034, P.R. China \\
3) School of Nuclear Science and Technology, Lanzhou University,
Lanzhou 730000, P.R. China \\
4) China Institute of Atomic Energy, Beijing 102413, P.R. China\\
5) GSI Helmholtzzentrum f\"ur Schwerionenforschung GmbH, D-64291 Darmstadt, Germany\\
6) H. Niewodnicza\'nski Institute of Nuclear Physics, Pl-31342 Krak\'ow, Poland \\
\\
 }


\begin{abstract}
The Ultrarelativistic Quantum Molecular Dynamics (UrQMD) model, a microscopic
transport model, is used to study the directed and elliptic
collective flows and the nuclear stopping in Au+Au collisions at incident
energies covered by INDRA and lower-energy FOPI experiments. It is seen clearly
that these observables are sensitive to both, the potential terms (including
iso-scalar and iso-vector parts as well as the momentum dependent
term) in the equation of state (EoS) and the collision term
(including the Pauli-blocking and the medium-modified
nucleon-nucleon elastic cross section (NNECS)). The momentum
modifications of both, the mean-field potentials and the density
dependent NNECS, are found to be sensitive to the collectivity of
heavy-ion collisions. At INDRA energies ($\le 150$~MeV/nucleon),
the dynamic transport with
a soft EoS with momentum dependence and with the momentum-modified
density-dependent NNECS describes the directed flow exhibited by hydrogen isotopes
($Z=1$) emitted at mid-rapidity fairly well.
\end{abstract}

\keywords{Non-equilibrium dynamics, collective flow, nuclear stopping, transition energy}

\pacs{25.70.-z,24.10.-i,25.75.Ld} \maketitle

\section{Introduction}
Nuclear reactions at intermediate energies as, e.g., provided by the
heavy-ion synchrotron SIS at GSI have been investigated for several
decades but the physics governing their mechanisms are not yet
thoroughly understood and further investigations seem necessary.
Recently the two experimental collaborations INDRA and FOPI have
published a series of systematic observations made in experiments at
the GSI laboratory regarding several physical quantities such as the
collective flows, the nuclear stopping, and the light cluster and
produced-particle, i.e. mostly pion, production
\cite{Lukasik:2004df,Andronic:2006ra,Reisdorf:2006ie,Reisdorf:2010ab}.
Comparisons with model calculations (mainly performed with the
Quantum Molecular Dynamics (QMD) model), revealed some discrepancies
which deserve deeper investigations, mainly concerning dynamical
observables. The excitation functions of the directed and elliptic
collective flows and the nuclear stopping cannot be satisfactorily
described with the same equation of state (EoS) of nuclear matter
(see e.g. \cite{Danielewicz:2002pu,lukasik_unpublished}).

Important progress in nuclear physics at intermediate energies was
made recently by reaching consensus on the soft nature of the EoS of
nuclear matter
\cite{Magestro:2000ba,Li:2008gp,Fuchs:2007vt,Danielewicz:2002pu,Hartnack:2005tr}.
However, according to the theory of quantum hydrodynamics (QHD),
both the mean field and the two-body collisions have the same
origin, the effective Lagrangian density, based on which the medium
modifications of both the collisions and the mean-field transport
between the collisions should be considered self-consistently
\cite{Danielewicz:1982kk,Danielewicz:1982ca,Chou:1984es,Mao:1994zza,Li:2000sha}.
More explicitly, the medium-modified terms should take the density
dependence, the isospin asymmetry, and the momentum constraints into
account. In past investigations, these medium effects were
considered mainly in the mean-field part but usually not in the
collision part, leading to an obvious lack of self-consistency.
Hence, conclusions based on these treatments are not fully reliable,
and it is easy to understand why new difficulties arose in the
recent comparisons of data with model calculations.

In the lower range of SIS energies, the collision rate is known to
increase with beam energy and the interplay between mean field and
two-body collisions leads to a colorful phenomenology of the nuclear
dynamics as, e.g., apparent in the excitation functions of
collective
flows~\cite{Bertsch:1987ja,Lemmon:1999aa,Cussol:2002,Andronic:2006ra,Reisdorf:2006ie,Reisdorf:2010ab}:
with the increase of beam energy beyond the INDRA range ($\le
150$~MeV/nucleon), the slope, near mid-rapidity and with respect to
rapidity, of the directed flow of free protons or light clusters
changes sign from negative to positive, while the value of the
elliptic flow at mid-rapidity changes from positive (in-plane) to
negative (squeeze-out). These transitions of the directed and
elliptic flows imply clearly that the strength of two-body
collisions and the nuclear stopping power increase, which has been
observed in recent INDRA experiments as
well~\cite{Lukasik:2004df,Andronic:2006ra,Lehaut:2010}.
Theoretically, in addition to the well-known density-dependent
modifications of binary cross sections, the momentum constraints
will have to be considered seriously in order to describe these data
systematically~\cite{Mao:1994et,Li:2003vd,Sammarruca:2005ch,Sammarruca:2005tk,Li:2005jy,Li:1993rwa,Li:1993ef,Giansiracusa:1996zz,Li:2006ez,Zhang:2007gd,Danielewicz:2009eu,Li:2009px}.

In our previous work, we have attempted to build up a microscopic transport
model in which the dynamic process of heavy ion collisions (HICs) is
described in a more consistent and complete manner. The
Ultrarelativistic Quantum Molecular Dynamics (UrQMD) model is
adopted as a microscopic transport basis~\cite{Bass98,
Bleicher99,Bratkovskaya:2004kv}. In Refs.
\cite{Li:2005zza,LI:2005zi,Li:2005gfa,Li:2008bk}, both the
mean-field potentials and the relativistic effects on the relative
distance and momentum in potentials are treated comprehensively.
In Refs.~\cite{Li:2006ez,Li:2009px}, the medium modifications of the
nucleon-nucleon elastic cross sections (NNECS) are further taken
into account. With the so improved version of the UrQMD model, several
experimental observables at SIS and AGS energies were successfully
described~\cite{Li:2005gfa,Li:2008bk,Petersen:2006vm,Yuan:2010ad,Trautmann:2009kq,Trautmann:2010at}.
With the further consideration of the ``pre-formed'' hadron
potentials, the well-known HBT time-related puzzle throughout the
beam energy range from SIS up to RHIC is well understood~\cite{Li:2007yd,Li:2008ge}.
It is also found, however, that further investigation is still needed at the lower SIS
energies~\cite{Yuan:2010ad}. This will become evident in this work from the study of
collective flows and nuclear stopping.

The paper is arranged as follows: Section II presents the new
updates in the UrQMD transport model. Results for collective flows
and nuclear stopping in the energy range covered by INDRA and FOPI
experiments are then shown in Section III. Finally, a conclusion and
an outlook are given in Section IV.

\section{UrQMD transport model and updates}

The UrQMD model is based on analogous principles as the QMD model
\cite{Aichelin:1986wa,Aichelin:1991xy} and the Relativistic Quantum
Molecular Dynamics (RQMD) model \cite{Sorge:1989dy}. The first
formal version (ver1.0) of the UrQMD transport model was published
at the end of last century  \cite{Bass:1997xw,Bass98, Bleicher99}.
Since then, a large number of successful theoretical analyses,
predictions, and comparisons with data have been accomplished with
this transport model for {\it pp}, {\it p}A and AA reactions and over a
large range of beam energies, essentially from SIS over AGS, SPS, RHIC, up to
LHC energies~\cite{urqmdweb1}.

\subsection{Potential updates}

In the UrQMD, similar to the QMD, hadrons are represented by
Gaussian wave packets in phase space. After the initialization of
projectile and target nuclei for which the Hard-Sphere (H-S) or the
Woods-Saxon (W-S) mode may be selected, the phase space of hadron
$i$ is propagated according to Hamilton's equations of motion:

\begin{equation}
{\bf \dot{r}}_i=\frac{\partial H}{\partial {\bf p}_i},
\hspace{1.5cm} \mathrm{and} \hspace{1.5cm} {\bf
\dot{p}}_i=-\frac{\partial H}{\partial {\bf r}_i}. \label{Hemrp}
\end{equation}
Here ${\bf {r}}$ and ${\bf {p}}$ are the coordinate and momentum of
hadron $i$. The Hamiltonian $H$ consists of the kinetic energy $T$
and the effective two-body interaction potential energy $V$,

\begin{equation}
H=T+V, \label{Hamifunc}
\end{equation}
with
\begin{equation}
T=\sum_i (E_i-m_i)=\sum_i (\sqrt{m_i^2+{\bf p}_i^2}-m_i). \label{HT}
\end{equation}
In the standard version of the UrQMD model \cite{Bass:1997xw,Bass98,
Bleicher99}, the potential energies include the two-body and
three-body Skyrme-, and the Yukawa-, Coulomb-, and Pauli-terms as a
base,

\begin{equation}
V=V_{sky}^{(2)}+V_{sky}^{(3)}+V_{Yuk}+V_{Cou}+V_{Pau}. \label{HV}
\end{equation}
The three-body terms of the Skyrme force can be approximately written in the form of
two-body interactions, the so-called density dependent terms.
Only the Coulomb force between charged baryons is considered in the default version.
However, when also considered for charged mesons it was found indispensable for
reproducing certain observables~\cite{Li:2005zza,Li:2008bk}.

The single particle potential follows from $U= \delta V/\delta
f$, where $f$ is the phase-space Wigner distribution function which
reads as

\begin{equation}
f({\bf r},{\bf p})=\sum_i f_i({\bf r},{\bf p})=\sum_i \frac{1}{(\pi
\hbar)^3}e^{-({\bf r}-{\bf r}_i)^2/2L^2}e^{-({\bf p}-{\bf
p}_i)^2\cdot 2L^2/\hbar^2}, \label{wigfunc}
\end{equation}
where $L$ is the width parameter of the wave packet.  It is known
that the width affects the stability of initialized nuclei and
the production of clusters at later reaction stages \cite{Hartnack:1997ez}.
Empirically, the value of the wave packet width should grow
from about 1 fm to 2 fm as the total mass of the colliding system increases
from light to heavy. As an example, for Au+Au collisions a
value of $L = 2$ fm is chosen while for Ca+Ca and lighter nuclei $L = 1$
fm is found more appropriate.

Recently, in order to permit more successful applications in the
intermediate energy regime ($0.1A \lesssim E_{lab} \lesssim 2A$
GeV), more potential terms have been incorporated into the UrQMD~\cite{Li:2005gfa}.
The first addition is the density-dependent average symmetry potential energy term
$V_{sym}^{pot}\delta^2$, where $\delta=(\rho_n-\rho_p)/\rho_0$ is the isospin-asymmetry.
In this work, the form

\begin{equation}
V_{sym}^{pot}=(S_0-\frac{\epsilon_F}{3})u^\gamma
\label{sympot}
\end{equation}
introduced in Ref.~\cite{Li:2002qx} is used. Here $S_0$ is the
symmetry energy at the normal nuclear density $\rho_0$, $\epsilon_F$
is the Fermi kinetic energy at normal nuclear density,
$u=\rho/\rho_0$ is the reduced nuclear density, and $\gamma$ is the
strength of the density dependence of the symmetry potential. In
this work, the value $S_0=34$ MeV is chosen. The symmetry potential
is important for isospin-asymmetric reactions at intermediate and
low energies since the uncertainties existing in the symmetry energy
based on various theories were found to be large at densities away
from the normal density \cite{Brown:2000pd}. Here, a soft
($\gamma=0.5$) density dependent symmetry potential is adopted.
Recent investigations indicate that the density dependence of the
symmetry energy at subnormal densities might be soft
\cite{Li:2005jy,Famiano:2006rb,Tsang:2008fd}. However, this will
still have to be confirmed by further studies.

The second addition is the momentum-dependent term \cite{Bass:1995pj}
\begin{equation}
U_{md}=t_{md} \ln^2[1+a_{md}({\bf p}_i-{\bf p}_j)^2]\rho_i/\rho_0,
\label{umd}
\end{equation}
where $t_{md}=1.57$ MeV and $a_{md}=500$ c$^2/$GeV$^2$ are selected
as in previous calculations within the QMD family. It has to be
noted, however, that this form may also have to be refined in order
to obtain improved fits of the real part of the optical
potential~\cite{Isse:2005nk,Li:2008gp}.

At beam energies higher than the AGS regime, potentials for ``pre-formed''
particles (string fragments) and relativistic effects regarding the
relative distance and the relative momentum employed in the two-body
potentials (Lorentz transformation) are, furthermore, considered
\cite{Li:2007yd,Li:2008bk}. They can be neglected in this work here.

\subsection{New Pauli-blocking treatment for two-body collision}
Regarding the collision term, it is well-known that the rate of
collisions is reduced by the quantum Pauli-blocking. However, in the
default version of the UrQMD, for simplicity, the Pauli-blocking
criteria are realized with an algorithm based on fitted parameters:
for the same type of particles, the binary scattering will be
blocked under the form

\begin{equation}
\eta<\psi , \label{pbdef}
\end{equation}
where
\begin{equation}
\psi=\sum_{i}e^{-\frac{({\bf r}-{\bf r_i})^2}{4L^2}}e^{-\frac{({\bf
p}-{\bf p_i})^2 \cdot L^2}{\hbar^2}} ,\label{fbl}
\end{equation}
and

\begin{equation}
\eta=a_{fit}+b_{fit}\varrho \label{pbdefault1}
\end{equation}
with
\begin{equation}
 \varrho=\sum_{i}e^{-\frac{({\bf r}-{\bf r_i})^2}{2L^2}} .\label{rhobl}
\end{equation}
The parameters $a_{fit}$ and $b_{fit}$ in Eq.\ (\ref{pbdefault1})
are set to be 1.49641 and 0.208736, respectively. It will be shown
below that, for descriptions of experimental observables at the
lower energies, the default Pauli-blocking treatment is not precise
enough. Therefore, an updated new version of the Pauli-blocking
routine is introduced here. For the same type of particles and for
each collision, firstly, the phase space densities in the final
states are determined in order to assure that they are in agreement
with the Pauli principle and, secondly, the following two criteria
are considered at the same time:

\begin{equation}
\frac{4\pi}{3}r_{ij}^3\cdot \frac{4\pi}{3}p_{ij}^3 \geq (2s+1)\cdot
(\frac{h}{2})^3 \label{newbl1}
\end{equation}

and
\begin{equation}
P_{block}=1-(1-f_i)(1-f_j) < \xi . \label{newbl2}
\end{equation}
On the left-hand side of Eq.\ (\ref{newbl1}), the $r_{ij}$ and
$p_{ij}$ are the relative distance and momentum of the two particles
at the final states $i$ and $j$. The factor $(2s+1)$ on the right
side of Eq.\ (\ref{newbl1}) denotes the summation of the spins of
the two particles. In Eq.\ (\ref{newbl2}), $\xi$ represents a random
number between 0 and 1. If one of the criteria is not fulfilled the
collision is not allowed and the two particles remain with their
original momenta.

\subsection{Medium modifications and momentum dependence of NNECS}

It is also known that the cross sections will be modified by the
nuclear medium, according to the QHD theory (see, e.g., Refs.\
\cite{Chou:1984es,Brockmann:1990cn,Li:1993rwa,Li:1993ef,Mao:1994zza,Hofmann:2000vz,Li:2000sha,Li:2003vd}.
As in previous work, the in-medium NNECS $\sigma_{el}^{*}$ are
treated to be factorized as the product of a medium correction
factor $F(u,\alpha,p)$ and the free NNECS
$\sigma_{el}^{free}$~\cite{Li:2006ez}. For the inelastic channels,
we still use the experimental free-space cross sections
$\sigma_{in}^{free}$. It is believed that this assumption does not
have strong influence on our present study at low SIS energies.
Therefore, the total two-body scattering cross section of nucleons,
$\sigma_{tot}^{*}$, will be modified according to
\begin{equation}
\sigma_{tot}^{*}=\sigma_{in}+\sigma_{el}^{*}=\sigma_{in}^{free}+F(u,\alpha,p)
\sigma_{el}^{free}.
\label{ecsf}
\end{equation}
As for the medium correction factor $F(u,\alpha,p)$, it is
proportional to both the isospin-scalar density effect $F_u$ and the
isospin-vector mass-splitting effect $F_\alpha$. Studies of the
isospin-related splitting effect on NNECS which is represented by
the $F_\alpha$ factor have been reported in Refs.\
\cite{Li:2006ez,Li:2009px}.

\begin{table}
\begin{center}
\renewcommand{\arraystretch}{1.2}
\begin{tabular}{|l|c|c|}\hline
\bf Set & $\lambda$ & $\zeta$ \\\hline\hline
\tt FU1 \ \ &\  1/3 \  & \ 0.54568  \    \\
\tt FU2 \ \ &\  1/4 \  & \ 0.54568  \      \\
\tt FU3 \ \  &\  1/6 \  & \ 1/3    \    \\ \hline
\end{tabular}
\end{center}
\caption{The three parameter sets FU1, FU2, and FU3 used for the
density-dependent correction factor $F_u$ of NNECS.} \label{tabfu}
\end{table}

\begin{table}
\begin{center}
\renewcommand{\arraystretch}{1.2}
\begin{tabular}{|l|c|c|c|}\hline
\bf Set & $f_0$ & $p_0$ [GeV c$^{-1}$]   & $\kappa$  \\\hline\hline
\tt FP1 & 1   & 0.425 & \ 5  \        \\
\tt FP2 & 1   & 0.225 & 3        \\
\tt FP3 & 1   & 0.625 & 8        \\ \hline
\tt no $p_{NN}$ limit & $F(u)$ & / &/  \\ \hline
\end{tabular}
\end{center}
\caption{The three parameter sets FP1, FP2, and FP3 used for
describing the momentum dependence of $F_u$. The fourth case,
without a $p_{NN}$ limit, is obtained by setting $f_0$ equal to
$F(u)$ in Eq.\ (\ref{fdpup}).} \label{tabfp}
\end{table}

\begin{figure}
\includegraphics[angle=0,width=0.8\textwidth]{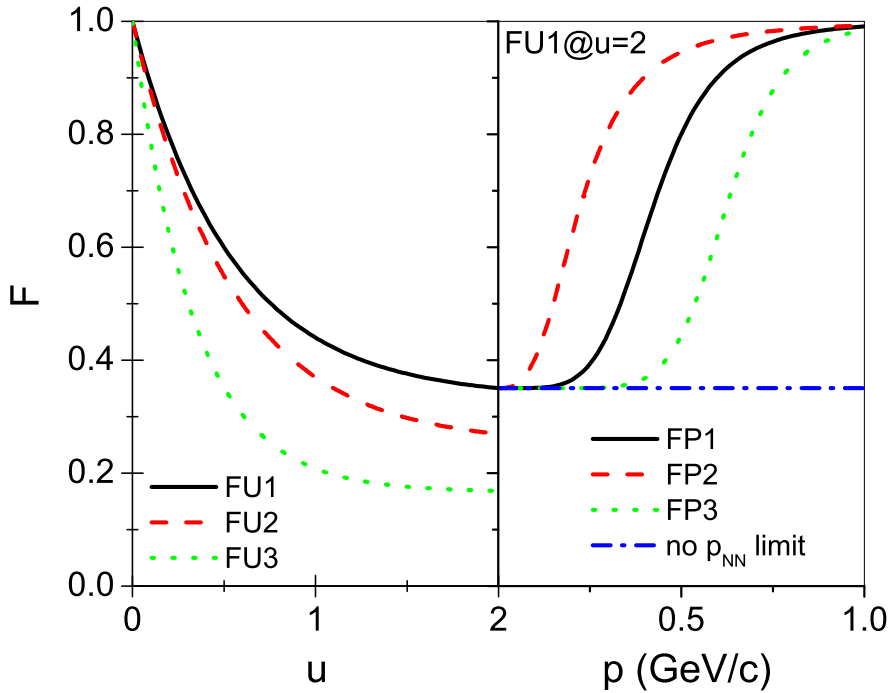}
\caption{(Color online) Correction factor $F$ obtained with the
parameterizations FU1, FU2, and FU3 given in Table \ref{tabfu} (left
panel) and the momentum dependence obtained with the four options
FP1, FP2, FP3, and ``no p$_{NN}$ limit'' given in Table \ref{tabfp}
for the example of FU1 at $u=2$ (right panel).} \label{fig1}
\end{figure}

Furthermore, the factors $F_u$ and $F_\alpha$ should be functions
of the relative momentum $p_{NN}$ of the two colliding particles in the NN
center-of-mass system.
In Ref.~\cite{Li:2006ez}, they are formulated as

\begin{equation}
F_{\alpha,u}^{\rm p}=\left\{
\begin{array}{l}
f_0 \hspace{3.45cm} p_{NN}>1 {\rm GeV}/c \\
\frac{F_{\alpha,u} -f_0}{1+(p_{NN}/p_0)^\kappa}+f_0 \hspace{1cm}
p_{NN} \leq 1 {\rm GeV}/c.
\end{array}
\right.
\label{fdpup}
\end{equation}
The factor $F_u$ can be expressed as
\begin{equation}
F_u=\lambda+(1-\lambda)\exp[-u/\zeta]. \label{fr}
\end{equation}
Here $\zeta$ and $\lambda$ are parameters which determine the
density dependence of the cross sections. In this work, we select
several parameter sets which are shown in Table \ref{tabfu} and
illustrated in the left panel of Fig.\ \ref{fig1}. The reduction of
the NNECS as a function of density becomes increasingly more
pronounced as the parameterization is changed from FU1 to FU3. At
the reduced density $u=2$, e.g., the values of FU1, FU2, and FU3 are
0.35, 0.27, and 0.17, respectively. We note here that the density
dependence of the FU1 parameterization is in qualitative agreement
with previous work based on the Dirac-Brueckner approach
\cite{Li:1993rwa,Li:1993ef,Fuchs:2001fp}. However, in our previous
investigations of the NNECS, based on the effective Lagrangian of
density-dependent relativistic hadron theory in which the $\sigma$,
$\omega$, $\rho$ and $\delta~[a_0(980)]$ mesons are included
\cite{Li:2003vd}, it was shown that especially the neutron-proton
cross sections $\sigma_{el,np}^*$ might be largely reduced in the
neutron-rich nuclear medium; the corresponding reduction factor
might be as low as $\sim 0.1$ at $u=2$. Therefore, the other
parameter sets FU2 and FU3 (Table \ref{tabfu}) are still to be
considered reasonable assumptions.

The parameters $f_0$, $p_0$ and $\kappa$ in Eq.\ (\ref{fdpup}) can
be varied in order to obtain various momentum dependences of $F_u$.
In this work, we select several parameter sets which are shown in
Table~\ref{tabfp}. The corresponding $F_u^p$ functions are
illustrated in the right panel of Fig.\ \ref{fig1} for FU1 at a
reduced density $u=2$. The FP1 set was used in our previous work in
which the medium modifications of cross sections were
considered~\cite{Li:2006ez}. The parameterizations FP2 or FP3 result
in a rise of the correction factor $F$ at smaller or at larger
$p_{NN}$ than obtained with FP1. It illustrates the present
uncertainty associated with treating the momentum dependence of the
density-dependent cross sections. With a certain set of isospin
dependent EoS, the NNECS might even be enhanced at large momenta, in
comparison to the cross sections at free space. It arises from the
differences between the isoscalar and isovector
channels\cite{Li:2000sha}. An indication of this enhancement was
obtained from a recent calculation of flow observables by Zhang {\it
et al.}~\cite{Zhang:2007gd}. But it will not be discussed in the
current work since the effect is related to collisions at large
$p_{NN}$ and it becomes increasingly important for heavy-ion
collisions at higher beam energies. And, for completeness, the case
without any momentum constraint on $F_u$ (``no p$_{NN}$ limit'') is
listed in Table~\ref{tabfp}.

\section{Calculations and Observables}

\subsection{Colliding system and the after-burner}

About 120 thousand events of $^{197}$Au+$^{197}$Au collisions for
each of the energies $E_{lab}=40$, $50$, $60$, $80$, $100$, $150$,
and $400$~MeV/nucleon are calculated randomly within the impact
parameter region 0-7.5 fm. For each beam energy, the calculations
are divided into three groups according to impact parameter
representing the central (0-2 fm), the semi-central (2-5.5 fm), and
the semi-peripheral (5.5-7.5 fm) group of collisions. Fourteen
parameterizations of the UrQMD transport model differing in the
treatments of the Pauli-blocking of two-body collisions (P-B), of
the potential terms (EoS), of the medium-modified NNECS with the
density-dependent factor $F_u$ and the momentum modification $F^p$
of $F_u$, and of the initialization (ini.) are shown in
Table~\ref{tab3} and will be discussed in the following section. The
UrQMD transport program stops and produces the output after a
collision time of 150 fm$/c$. Several more outputs are produced, as
the calculations evolve, at different reaction times before the stop
time. For each output, a conventional phase-space coalescence
model~\cite{Kruse:1985pg} with two parameters is used to construct
clusters. Nucleons with relative momenta smaller than $P_0$ and
relative distances smaller than $R_0$ are considered to belong to
one cluster. In this work, $P_0$ and $R_0$ are chosen to be 0.2
GeV$/c$ and 2.8 fm, respectively. It is evident and well-known that
the criteria chosen for constructing clusters affect their yields
and, consequently, also related observables~\cite{Li:2008fn}.

\begin{table}
\begin{center}
\renewcommand{\arraystretch}{1.2}
\begin{tabular}{|l|l|l|l|l|l|}\hline
\bf Set        &   P-B  &  EoS            &  $F_u$   &  $F^p$   &
ini. \\\hline\hline
\tt UrQMD-0    &   default  &  SM             &  FU1     &  FP1     &  H-S  \\
\tt UrQMD-I    &   new  &  SM             &  FU1     &  FP2     &  H-S  \\
\tt UrQMD-II   &   new  &  SM             &  FU1     &  FP1     &  H-S  \\
\tt UrQMD-III  &   new  &  SM             &  FU1     &  FP1     &  W-S  \\
\tt UrQMD-IV   &   new  &  Cascade        &  FU1     &  FP1     &  W-S  \\
\tt UrQMD-V    &   new  &  SM(no $V_{sym}$+$V_{Cou})$ &  FU1     &  FP1     &  H-S  \\
\tt UrQMD-VI   &   new  &  SM             &  no Collisions &  no Collisions &  H-S  \\
\tt UrQMD-VII  &   new  &  SM             &  FU1     &  FP3     &  H-S  \\
\tt UrQMD-VIII &   new  &  SM             &  FU1     &  no $p_{NN}$ limit       &  H-S  \\
\tt UrQMD-IX   &   new  &  S              &  FU1     &  FP1     &  H-S  \\
\tt UrQMD-X    &   new  &  S              &  FU1     &  no $p_{NN}$ limit       &  H-S  \\
\tt UrQMD-XI   &   new  &  S              &  FU2     &  no $p_{NN}$ limit       &  H-S  \\
\tt UrQMD-XII  &   new  &  S              &  FU3     &  no $p_{NN}$ limit       &  H-S  \\
\tt UrQMD-XIII  &   new  &  SM             &  FU3     &  FP1     &  H-S  \\
\hline
\end{tabular}
\end{center}
\caption{Fourteen parameterizations of the UrQMD transport model
differing in the treatments of the Pauli-blocking of two-body
collisions (P-B), of the potential terms (EoS), of the
medium-modified NNECS with the density-dependent factor $F_u$ and
the momentum modification $F^p$ of $F_u$, and of the initialization
(ini.). See text for details.} \label{tab3}
\end{table}

\subsection{Collective flows and the stopping $vartl$}

In this work, we will focus on the parameters describing the strength and orientation of
directed and elliptic collective flows
defined as

\begin{equation}
v_1\equiv <cos(\phi-\Phi_{RP})>=<\frac{p_x}{p_t}>, \label{eqv1}
\end{equation}
and
\begin{equation}
v_2\equiv <cos[2(\phi-\Phi_{RP})]>=<\frac{p_x^2-p_y^2}{p_t^2}>.
\label{eqv2}
\end{equation}
Here $\phi$ denotes the azimuthal angle of the considered outgoing particle;
$\Psi_{RP}$ is the azimuthal angle of the reaction plane which is pre-configured
in the calculations, $\Psi_{RP}=0$, but has to be determined in experiments;
$p_x$ and $p_y$ are the two components of the transverse
momentum $p_t=\sqrt{p_x^2+p_y^2}$. The angular brackets denote an
average over all considered particles from all events.

The nuclear stopping is described with the quantity
$vartl$~\cite{Reisdorf:2004wg} defined as

\begin{equation}
vartl=\frac{\Gamma_{dN/dy_x}}{\Gamma_{dN/dy_z}} . \label{eqvartl}
\end{equation}
Here
\begin{equation}
\Gamma_{dN/dy_{x,z}}=<y_{x,z}^2>=\frac{\sum(y^2_{x,z}N_{y_{x,z}})}{\sum
N_{y_{x,z}}}, \label{eqgm}
\end{equation}
where $\Gamma_{dN/dy_{x}}$ and $\Gamma_{dN/dy_{z}}$ are the
variances of the rapidity distributions of fragments in the $x$ and
$z$ directions, respectively. They are obtained as weighted averages
of $y^2$ over the considered ranges in rapidity with $N_{y_{x}}$ and
$N_{y_{z}}$ denoting the yields of fragments in each of the $y_x$
and $y_z$ rapidity bins.

Directed  and elliptic flows and the nuclear stopping quantity
$vartl$ were measured by the INDRA and FOPI collaborations at
GSI/Darmstadt
\cite{Lukasik:2004df,Andronic:2006ra,Reisdorf:2006ie,Reisdorf:2010ab}.
In the following, the predictions of the UrQMD model for these
observables will be analyzed, while the directed flow is chosen for
the comparison with the experimental data at INDRA energies
($E_{lab}\leq 150$~MeV/nucleon).

\section{Results}

\subsection{Pauli-Blocking effects on flows}

\begin{figure}
\includegraphics[angle=0,width=0.8\textwidth]{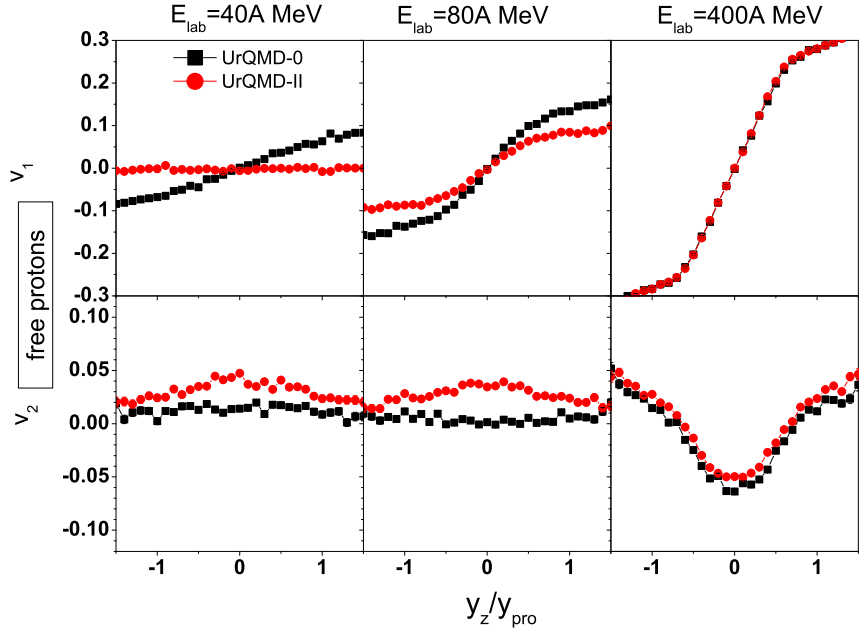}
\caption{(Color online) Parameters $v_1$ of directed flow (top
panels) and $v_2$ of elliptic flow (bottom panels) for free protons
from semi-peripheral ($b=5.5--7.5$~fm) $^{197}$Au+$^{197}$Au
collisions at $E_{\rm lab}=40$ (left), 80 (middle), and 400
MeV/nucleon (right), calculated with the versions UrQMD-0 (squares)
and UrQMD-II (circles), as a function of the reduced rapidity
$y_z/y_{\rm pro}$. } \label{fig2}
\end{figure}

As stated before, the Pauli-blocking treated in the default UrQMD
model is an effective one. The advantage of this treatment is the
high-speed to calculate it. However, it is found that it is not
quite suitable for nuclear reactions at low SIS energies. In
Fig.~\ref{fig2} we present the reduced longitudinal rapidity
distribution of directed ($v_1$, top plots) and elliptic ($v_2$,
bottom plots) flows of free protons from semi-peripheral (b=5.5-7.5
fm) Au+Au collisions at $E_{lab}=40$, 80, and 400 MeV/nucleon. The
calculations with the default treatment of Pauli-blocking used in
UrQMD-0 and with the new treatment incorporated in UrQMD-II (the
only difference between the two versions, cf. Table~\ref{tab3}) are
shown in Fig.~\ref{fig2}.

Considerable differences exist in the rapidity distribution of both
flow parameters at the lower SIS energies $E_{\rm lab}=40$ and 80
MeV/nucleon. With the default treatment, the slope of $v_1$ is
significantly positive at mid-rapidity and the value of $v_2$ is
rather small. Experimentally it is known that the slope of the
directed flow of $Z=1$ particles from semi-central/peripheral
collisions of heavy systems at 40 MeV/nucleon is slightly negative
at mid-rapidity and that the elliptic flow is positive with values
near $v_2 =0.08$~\cite{Lukasik:2004df}, values much closer to the
results obtained with the new treatment of Pauli-blocking
(Fig.~\ref{fig2}, left panels). It is, therefore, concluded that
this modification of the UrQMD is more suitable for this energy
range, and the new treatment of Pauli-blocking is chosen for the
following investigations. The differences start to disappear at the
higher energy 400 MeV/nucleon (right panels of Fig.~\ref{fig2}).

\subsection{Causes of collective phenomena}

\begin{figure}
\includegraphics[angle=0,width=0.8\textwidth]{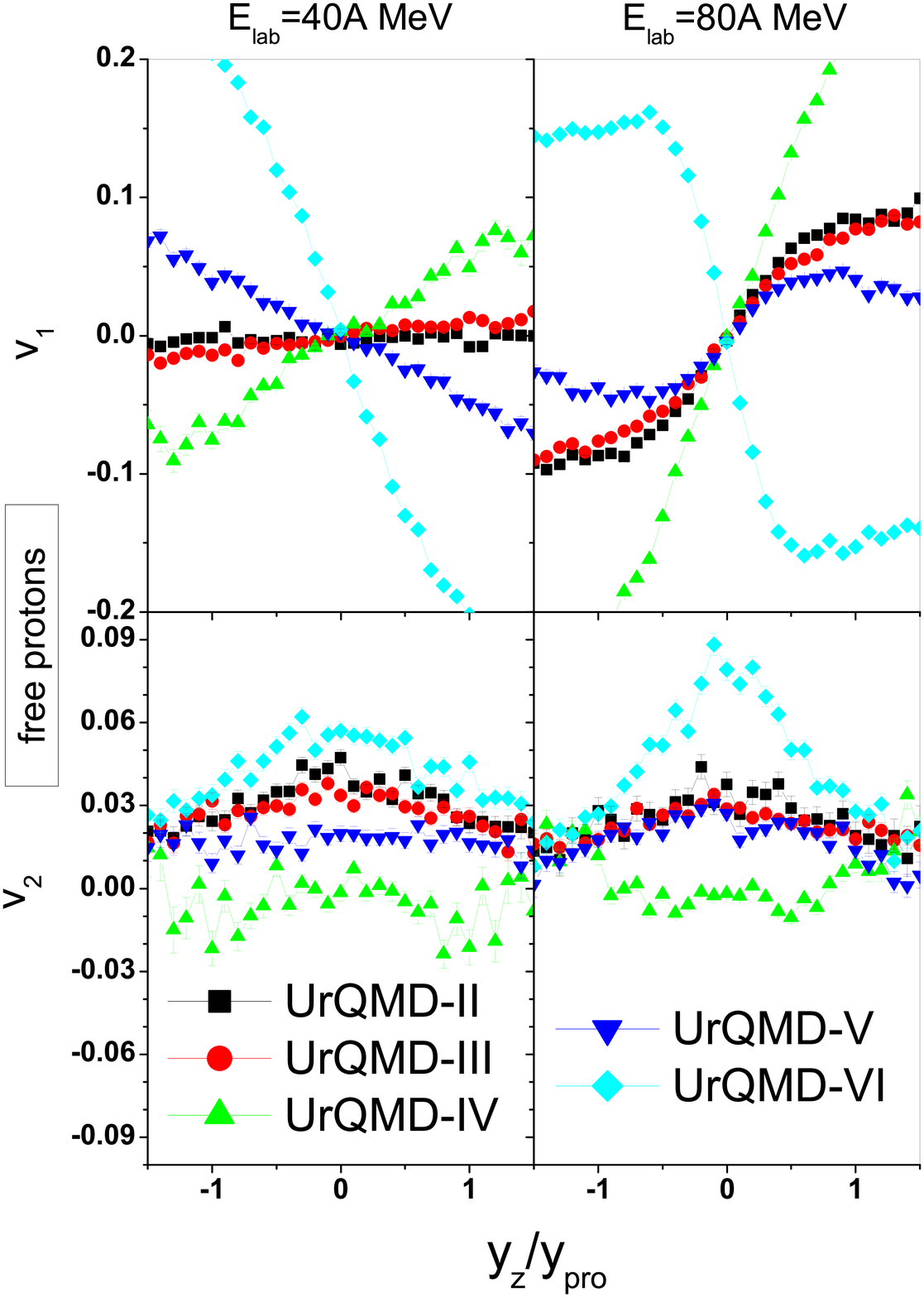}
\caption{(Color online) Parameters $v_1$ of directed flow (top
panels) and $v_2$ of elliptic flow (bottom panels) for free protons
from semi-peripheral ($b=5.5-7.5$~fm) $^{197}$Au+$^{197}$Au
collisions at $E_{\rm lab}=40$ (left) and 80 MeV/nucleon (right) as
a function of the reduced rapidity $y_z/y_{\rm pro}$. The
calculations performed with versions UrQMD-II, UrQMD-III, UrQMD-IV,
UrQMD-V, and UrQMD-VI (see Table \ref{tab3}) are distinguished by
their symbols as indicated.} \label{fig3}
\end{figure}

In order to understand how the initialization, the mean-field
potentials and collisions influence collective flows, calculations
were performed with the versions UrQMD-II, UrQMD-III, UrQMD-IV,
UrQMD-V, and UrQMD-VI (Fig.\ \ref{fig3}). Several interesting
phenomena are observed: comparing first the results obtained with
UrQMD-II and UrQMD-III which differ only in the initialization,
small differences are seen but the directed flow is slightly less
positive at 40 MeV/nucleon and slightly more positive at 80
MeV/nucleon with the Hard-Sphere treatment of UrQMD-II than with the
Wood-Saxon initialization of UrQMD-III. This is due to the fact that
at 40 (80) MeV/nucleon the net-contribution of both mean-field
potentials and two-body collisions is more attractive (positive) and
the Hard-Sphere initialization may lead to stronger attraction
(repulsion). Actually, this effect of different initializations was
seen before in calculations with the IQMD model
\cite{Hartnack:1997ez}. Since its effect on final flows is very
small, mainly the Hard-Sphere initialization will be used in the
following calculations. It is the default initialization in the
UrQMD model.

Secondly, if the mean-field potentials are ignored as in the
cascade-type version UrQMD-IV, a large positive slope of $v_1$ is
seen at both beam energies. It implies that the net contribution of
the collision term to the directed flow is repulsive. The elliptic
flow has almost vanished which should be due to both, a relatively
weak contribution of the collision term and the almost isotropic
differential NNECS. On the other hand, if the collision term is
switched off and only the mean-field potentials are active as in
UrQMD-VI, a strong negative directed flow is seen at both beam
energies. The elliptic flow $v_2$ is largest in this case,
indicating that the mean-field potentials contribute to the
collective flows much more strongly than the collision term at lower
SIS energies, a fact that is well-known. It is also found that the
net contributions of the mean-field and of two-body collisions to
the directed flow are attractive and repulsive, respectively. The
positive elliptic flow induced by the mean-field is caused by the
in-plane geometry at early reaction times.

The isospin-scalar and isospin-vector related parts of the
potentials may yield different contributions to the final flows.
This was tested with the results of UrQMD-II, UrQMD-IV, and UrQMD-V
in which a soft-EoS with momentum dependence (SM-EoS, the
corresponding incompressibility $K=200$ MeV) is considered, not
considered, and partially considered, respectively, while the same
collision term is adopted. It is found that the iso-scalar part of
the potentials produces a strong attractive force (comparison of
UrQMD-IV and UrQMD-V) while the iso-vector related part gives a
relatively weak repulsive force to the directed flow (see from
UrQMD-V to UrQMD-II). The repulsion caused by the iso-vector
potentials (symmetry and Coulomb potentials) is obvious since the
Coulomb potential of protons is always repulsive and much stronger
than the symmetry potential, regardless of whether the total effect
of the symmetry term is attractive or repulsive. The latter depends
on the isospin-asymmetry of the nuclear medium during the whole
dynamic process of HICs.

Summarizing this analysis, one finds clearly that 1) the balance of
contributions between mean-field potentials and the collision term
determines the vanishing of the directed flow at low SIS energies
with roughly equal strength; 2) in the mean-field potentials, the
iso-scalar part behaves as an attractive effect which is due to the
overall rather low densities of heavy-ion collisions at such low
beam energies. The iso-vector part is repulsive which is mainly due
to the mutual Coulomb repulsion of the protons.

\begin{figure}
\includegraphics[angle=0,width=0.75\textwidth]{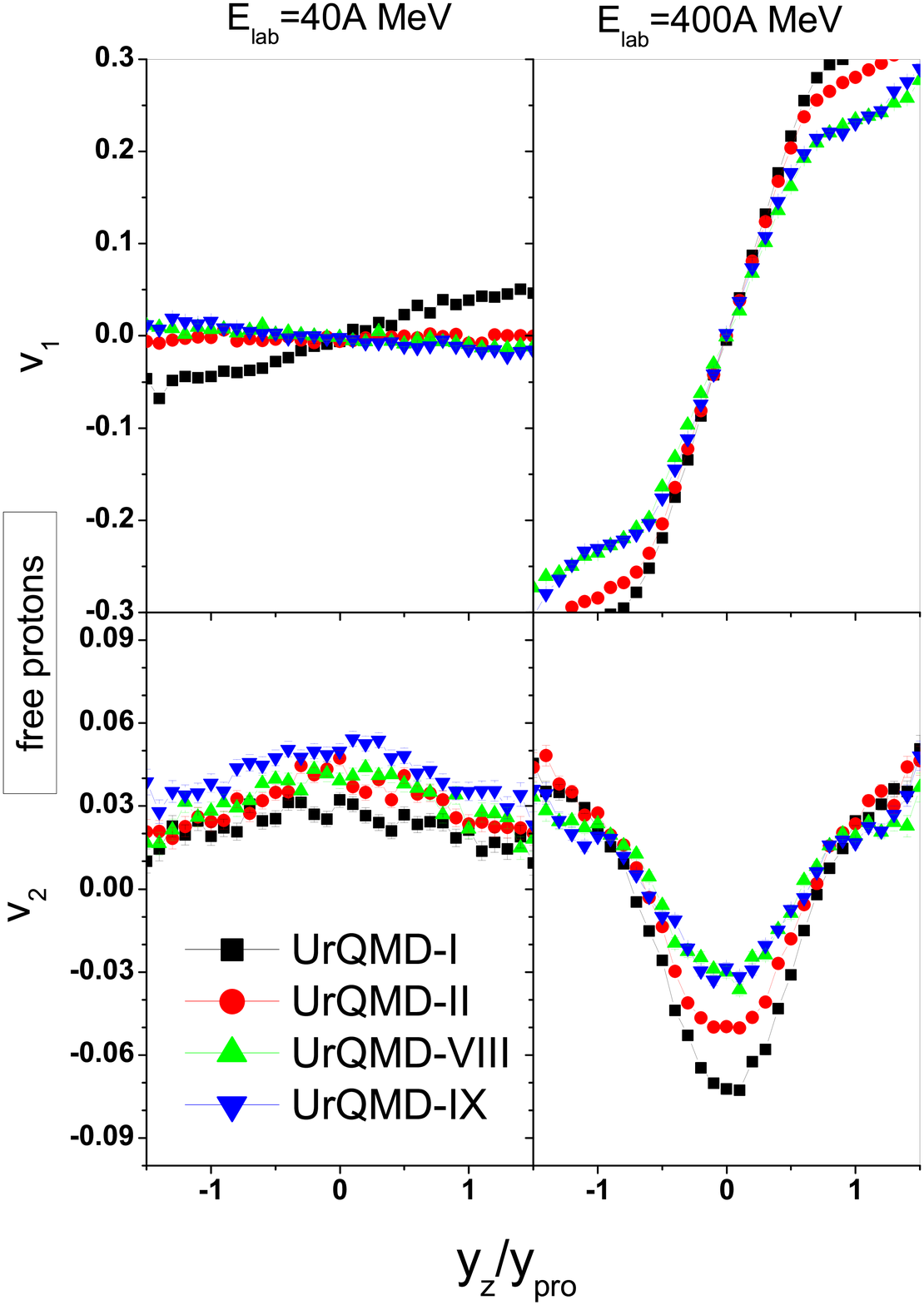}
\caption{(Color online) Parameters $v_1$ of directed flow (top
panels) and $v_2$ of elliptic flow (bottom panels) for free protons
from semi-peripheral ($b=5.5-7.5$~fm) $^{197}$Au+$^{197}$Au
collisions at $E_{\rm lab}=40$ (left) and 400~MeV/nucleon (right) as
functions of the reduced rapidity $y_z/y_{\rm pro}$. The
calculations performed with versions UrQMD-I, UrQMD-II, UrQMD-VIII,
and UrQMD-IX (see Table \ref{tab3}) are distinguished by their
symbols as indicated.} \label{fig4}
\end{figure}

Another set of comparisons of flow results, calculated with versions
UrQMD-I, UrQMD-II, UrQMD-VIII, and UrQMD-IX is shown in Fig.\
\ref{fig4}. Two beam energies 40 and 400~MeV/nucleon are selected.
The comparison of UrQMD-I, UrQMD-II, UrQMD-VIII shows the effects on
the flow parameters resulting from different momentum-modified NNECS
while the comparison of UrQMD-II and UrQMD-IX shows those of the
momentum-dependent terms in the potentials (with or without momentum
dependence). It is, first of all, noticed that both effects, the
momentum modification of the density-dependent NNECS in the
collision term and the momentum dependence of the potentials affect
both flow variables $v_1$ and $v_2$. Secondly, when studied in more
detail, these flow effects are found to behave differently at
different beam energies: (1) regarding the contribution of
medium-modified cross sections, we see that for the directed flow at
40~MeV/nucleon the difference is large between the UrQMD-I and
UrQMD-II results, while at 400~MeV/nucleon the largest difference
exists between the UrQMD-I and UrQMD-VIII results. It is obviously
caused by the difference of the momentum modifications of cross
sections (FP1, FP2, and no $P_{NN}$ limit) as they appear at
different energies; (2) for the elliptic flow at 40~MeV/nucleon, a
difference is seen between the UrQMD-VIII and UrQMD-IX results,
while this is no longer the case at 400~MeV. It implies that, at
40~MeV/nucleon,  the momentum-dependent term in the potentials is
more important than the momentum modification of the density
dependent NNECS. They become equally important at 400~MeV/nucleon.

A rough comparison of the calculations with INDRA and FOPI data for $Z=1$ particles
from the same system shows that the effect of the momentum-dependent terms
in the potentials is repulsive, i.e. not suitable for describing the measured
negative slope of the $v_1$ flow and the largely positive $v_2$ (in-plane flow)
at mid-rapidity and at the beam energy 40~MeV/nucleon. At
400~MeV/nucleon, the consideration of the momentum dependence in both, the
mean-field and collision terms, is important to describe the measured
largely positive slope of $v_1$ and largely negative $v_2$ (squeeze-out)
at mid-rapidity. Therefore, the uncertainties existing in the
momentum modification of NNECS will largely influence the
excitation functions of flows, even only at low SIS energies.

\begin{figure}
\includegraphics[angle=0,width=0.8\textwidth]{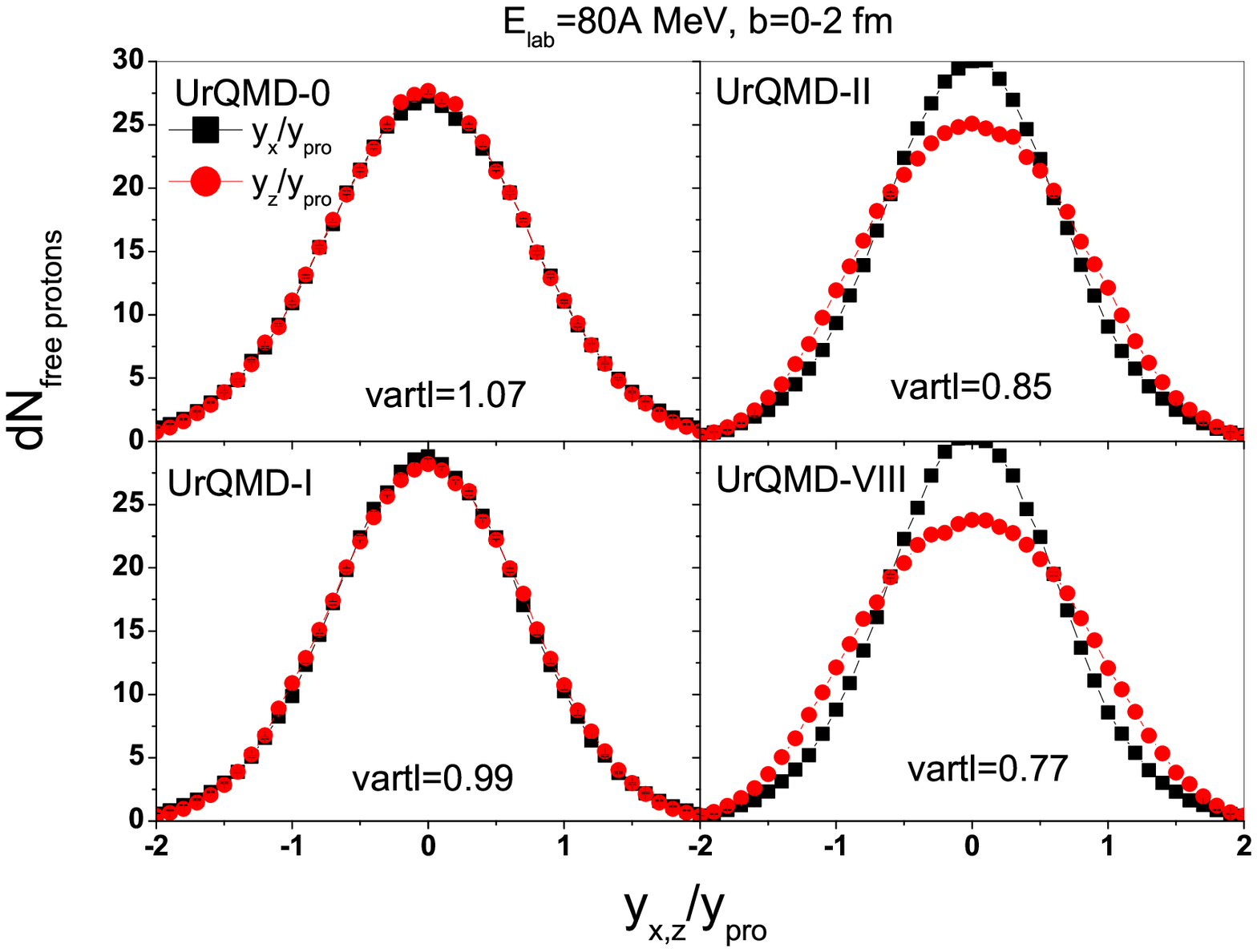}
\caption{(Color online) Yield distributions of free protons as
functions of the reduced longitudinal ($y_z/y_{\rm pro}$, squares)
and transverse ($y_x/y_{\rm pro}$, circles) rapidities from central
($b=0-2$~fm) $^{197}$Au+$^{197}$Au collisions at $E_{\rm
lab}=80$~MeV/nucleon. Calculations with UrQMD-0, UrQMD-I, UrQMD-II,
and UrQMD-VIII (see Table \ref{tab3}) are shown in the top-left,
bottom-left, top-right, and bottom-right panels, respectively. The
corresponding values obtained for the stopping observable $vartl$
for free protons are also indicated in each panel.} \label{fig5}
\end{figure}

The yield distributions of free protons as functions of the reduced
longitudinal ($y_z/y_{\rm pro}$) and transverse ($y_x/y_{\rm pro}$)
rapidities from central Au+Au collisions at $E_{\rm
lab}=80$~MeV/nucleon are shown in Fig.\ \ref{fig5}. Calculations
with the default Pauli-blocking treatment in UrQMD-0 (top-left) are
compared to those with the new Pauli-blocking treatment combined
with the same (UrQMD-II, top-right) or alternative momentum
modifications of NNECS (UrQMD-I, bottom-left and UrQMD-VIII,
bottom-right). The corresponding values of $vartl$, for the full
range of rapidities, are also given for each of the UrQMD versions.
We first notice that, in the UrQMD-0 mode, the relative maximum of
the proton yield is higher at mid-longitudinal-rapidity than at
mid-transverse-rapidity, which leads to $vartl>1$.  In contrast,
calculations with the new treatment of the Pauli-blocking, despite
of uncertainties in the cross sections, the values of $vartl$ are
always smaller than unity, in line with the experimental data.

Secondly, with decreasing cross sections (the reduction of NNECS due
to the momentum modifications increases successively going from
UrQMD-I to UrQMD-II and UrQMD-VIII), the calculated values of
$vartl$ become smaller and approach the experimental value $0.80\pm
0.03$ \cite{lukasik_unpublished} measured with INDRA at $E_{\rm
lab}=80$~MeV/nucleon (note, however, that one can not compare
directly with data since the experimentally selected clusters ($Z
\le 3$) and rapidity bins are different from the current
calculations performed for $Z=1$ particles). Together with the flow
results shown in Fig.\ \ref{fig4}, these comparisons clearly show
that the momentum modification of NNECS are important for describing
the collective flows and the nuclear stopping at the low SIS
energies at the same time. Thirdly, in the analyses related to
Figs.\ \ref{fig3}, \ref{fig4}, and \ref{fig5}, one finds that the
results for flows and for the nuclear stopping always follow in the
same order when different treatments of the mean-field and the
collisions terms are chosen. It, therefore, seems sufficient to
select one of the observables, the slope of the directed flow at
mid-rapidity, for the following more precise comparison with
experimental data.

\subsection{Comparison of the excitation function of the $v_1$ slope with INDRA data}

\begin{figure}
\includegraphics[angle=0,width=0.8\textwidth]{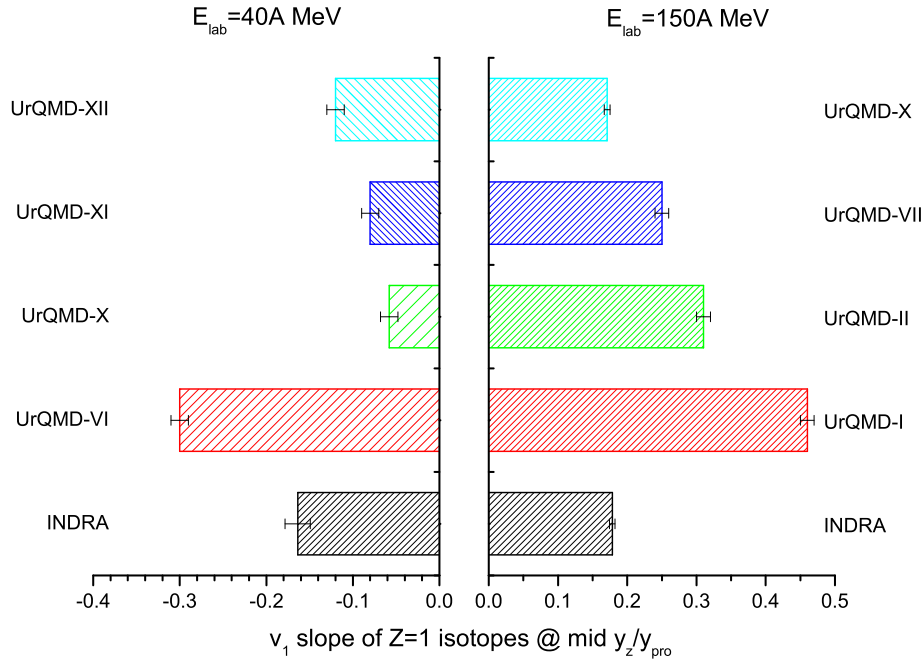}
\caption{(Color online) Values of the slope of directed flow at
mid-rapidity ($|y_z/y_{\rm pro}|<0.4$) for $Z=1$ particles from
semi-central ($b=2-5.5$~fm) $^{197}$Au+$^{197}$Au collisions at
$E_{lab}=40$~MeV/nucleon (negative values, left) and
$150$~MeV/nucleon (positive values, right). The INDRA results at
both beam energies are compared with UrQMD calculations performed
with the indicated versions (see Table \ref{tab3}), all represented
by the bars with different shadings. } \label{fig6}
\end{figure}

A modification of the density dependence might be considered as an
additional alternative to the momentum modification of NNECS in
order to achieve their stronger reduction in heavy-ion collisions at
lower energies and a weaker reduction at higher energies. In the
left half of Fig.\ \ref{fig6}, we show values of the slope of
directed flow at mid-rapidity ($|y_z/y_{\rm pro}|<0.4$) for $Z=1$
particles from semi-central ($b=2-5.5$~fm) Au+Au collisions at
$E_{\rm lab}=40$~MeV/nucleon. Calculations with UrQMD-VI, UrQMD-X,
UrQMD-XI, and UrQMD-XII are compared to the INDRA data. When the
collision term is turned off (UrQMD-VI), the absolute value of the
$v_1$ slope is larger than the data, indicating that collisions
start to become important for heavy-ion collisions at such low beam
energies. When the collision term is present and larger reductions
of NNECS are considered by choosing UrQMD-X, UrQMD-XI, and UrQMD-XII
(with increasingly larger density dependence but without momentum
modification), the obtained $v_1$ slopes clearly approach the
measured INDRA value.

\begin{figure}
\includegraphics[angle=0,width=0.6\textwidth]{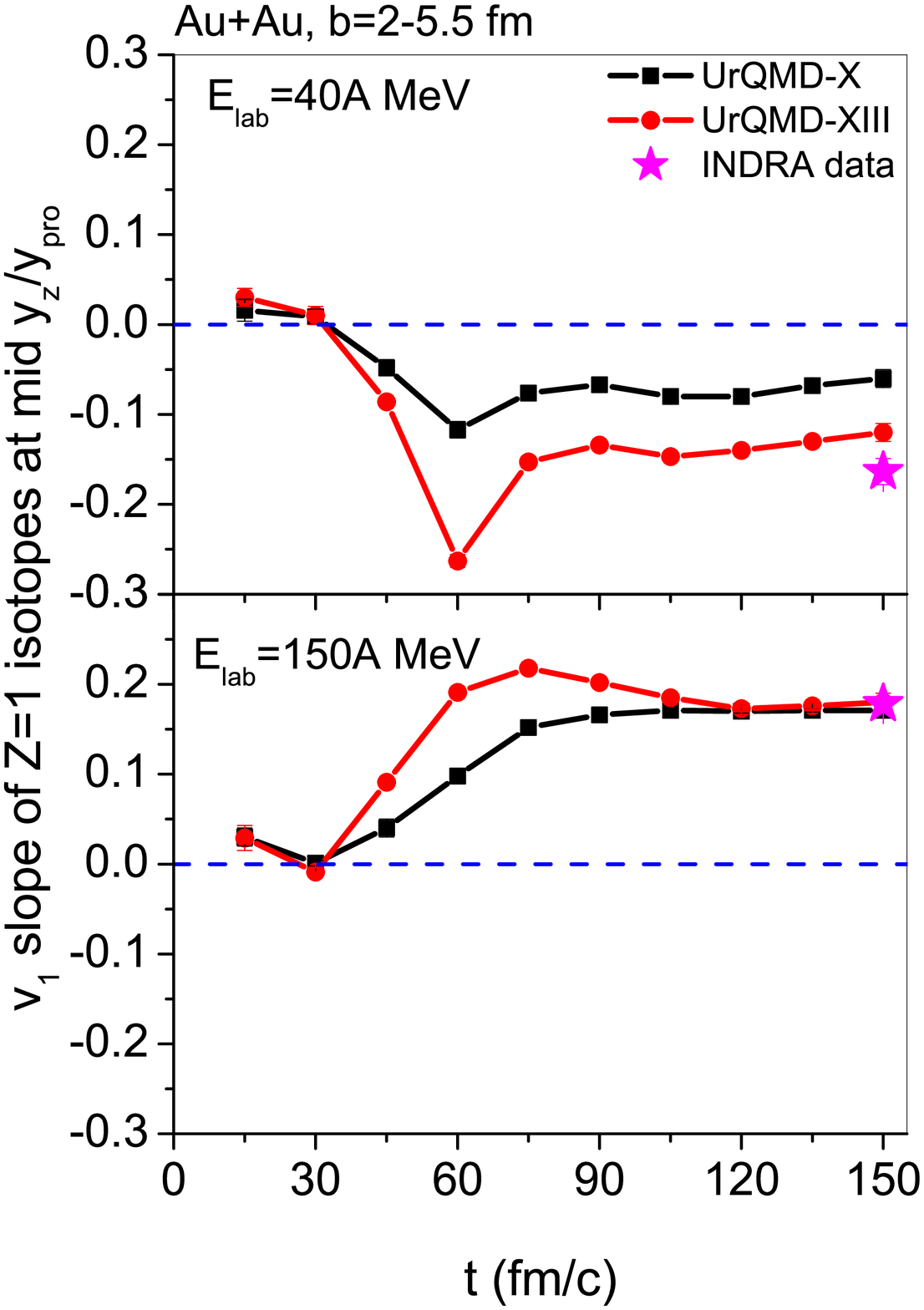}
\caption{(Color online) Time evolution of the slope of directed flow
at mid-rapidity ($|y_z/y_{\rm pro}|<0.4$) for $Z=1$ particles from
semi-central ($b=2-5.5$~fm) $^{197}$Au+$^{197}$Au collisions at
$E_{\rm lab}=40$~MeV/nucleon (top) and 150~MeV/nucleon (bottom).
Calculations with UrQMD-X and UrQMD-XIII (see Table \ref{tab3}) are
represented by the squares and circles, respectively. The INDRA
results for both beam energies are indicated by the stars positioned
at the UrQMD stop time of 150 fm/c. The dashed horizontal lines
representing zero slopes are included for comparison.} \label{fig7}
\end{figure}

On the right side of Fig.\ \ref{fig6}, again values of the slope of
directed flow at mid-rapidity for $Z=1$ particles from semi-central
Au+Au collisions but for an incident energy
$E_{lab}=150$~MeV/nucleon are presented. Calculations with UrQMD-I,
UrQMD-II, UrQMD-VII, and UrQMD-X are compared to the INDRA data.
Obviously, with the increasingly weaker momentum modification of
NNECS (as realized with UrQMD-I, UrQMD-II, and UrQMD-VII in that
sequence), the large reduction induced by the density dependence of
the NNECS is less compensated, leading to a weaker repulsion and,
hence, to the decreasing value of the $v_1$ slope shown in the
figure. Alternatively, if we switch off both the momentum-dependent
term in the potentials and the momentum modification of NNECS in the
collision term, realized with the UrQMD-X calculation, the value of
the $v_1$ slope is pushed down further and approaches the INDRA
data. As a result of these tests, one may reach a deeper
understanding of the difficulties in theoretically describing
experimental data and of the importance of treating
self-consistently the dynamic process of heavy-ion collisions.
Furthermore, it is evident from Fig.\ \ref{fig6} that even the
version UrQMD-X is still not good enough to permit the description
of the directed flow over the whole range of INDRA energies. Further
refinements and tests will be needed.

The time evolution of the slope of directed flow at mid-rapidity for
$Z=1$ particles from semi-central Au+Au collisions at $E_{lab}=40$
(top) and 150~MeV/nucleon (bottom) are shown in Fig.\ \ref{fig7}.
The versions chosen for the comparison are UrQMD-X, closest to the
INDRA data at 150~MeV/nucleon (Fig.~\ref{fig6}), and UrQMD-XIII. The
latter differs from UrQMD-XII, closest to the data at
40~MeV/nucleon, in that the SM-EoS and the FP1 momentum modification
of NNECS (the standard in previous investigations~\cite{Li:2006ez})
are adopted and from UrQMD-X by the stronger density dependence FU3
for NNECS (same as in UrQMD-XII). At about $t<30$~fm$/c$, the
pre-equilibrated protons are emitted with a small positive flow due
to the initial geometry. With increasing time, up to about
60~fm$/c$, the value of the $v_1$ slope increases or decreases,
depending on the balance between strong attractive and repulsive
effects on particles during the rescattering process. At
40~MeV/nucleon, the net contribution is attractive, whereas at
150~MeV/nucleon it is repulsive, leading to the corresponding
negative and positive flows. After 60~fm$/c$, the final-state
interactions (FSI) still affect the collisions at the low beam
energy 40~MeV/nucleon, and the contributions of two-body scatterings
start to become stronger than those of the mean field potentials. It
is found that, although the FSI affects the final flow results, the
UrQMD-X can not describe well the data at 40~MeV/nucleon. The
UrQMD-XIII calculations can describe reasonably well both results at
40 and 150~MeV/nucleon.

\begin{figure}
\includegraphics[angle=0,width=0.8\textwidth]{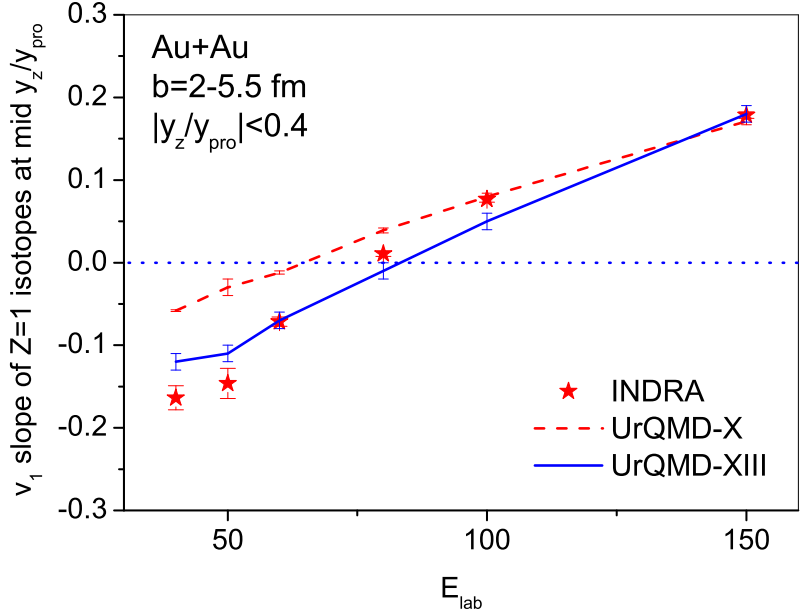}
\caption{(Color online) Excitation function of the slope of directed
flow at mid-rapidity ($|y_z/y_{\rm pro}|<0.4$) for $Z=1$ particles
from semi-central ($b=2-5.5$~fm) $^{197}$Au+$^{197}$Au collisions in
the range of INDRA energies from 40 to 150~MeV/nucleon. Calculations
with UrQMD-X (dashed) and UrQMD-XIII (full line) are compared with
the INDRA data (stars). The dotted horizontal line representing the
zero slope is included for comparison.} \label{fig8}
\end{figure}

Finally, in Fig.\ \ref{fig8}, the excitation function of
the slope of directed flow at mid-rapidity for $Z=1$ particles from semi-central
Au+Au collisions is shown for the range of INDRA energies from
40 to 150~MeV/nucleon. The data are from Ref.~\cite{Andronic:2006ra} and compared
to the results obtained with the same UrQMD-X and UrQMD-XIII whose time evolution has
just been discussed. It is seen clearly that the UrQMD-XIII calculations
describe the data well over the whole energy range including the location of
the transition energy which is closely reproduced. It has to be noticed, however,
that, in the energy range 90-150~MeV/nucleon covered also by FOPI experiments,
the FOPI result for the $v_1$~slope of $Z=1$ particles is higher by up to 0.05 than
the INDRA data~\cite{Andronic:2006ra}.
Thus, in order to pin down the exact form of the medium modifications
on both the mean-field potentials and the collision terms, it will also be necessary to
further reduce remaining uncertainties on the experimental side

\section{Conclusions and Outlook}

In summary, by using the microscopic transport model UrQMD, we have
performed a systematic investigation on the effects of the momentum
dependence of the mean field and the corrections from various
aspects in the collision term, especially the modification of the
density and momentum dependence of nucleon-nucleon elastic cross
sections (NNECS) on the observables, such as the directed and
elliptic flow and the stopping in HICs at low SIS energies. It is
clearly seen that the collective flows and the nuclear stopping are
sensitive to all these effects, meaning that they are important to
the non-equilibrium dynamic process in HICs at low SIS energies.

Further, in order to describe experimental observables
systematically, a consistent consideration of the uncertainties
associated with both, the mean-field and the two-body collisions, is
extremely important and should be paid more attention. The momentum
dependence of the mean field potentials and of the density-dependent
NNECS is found to be sensitive to the collectivity exhibited by the
collision dynamics of HICs. At INDRA energies (40-150~MeV/nucleon),
the dynamic transport with a soft equation-of-state with momentum
dependence (SM-EoS) and with momentum-dependent and density-modified
NNECS ($\sigma^*_{el}(\rho,p_{NN},\delta)$) describes the slope of
the directed flow of $Z=1$ particles at mid-rapidity rather well.

From Ref.\ \cite{Andronic:2006ra} we learned that the cluster
charge-Z dependence of the transition energy of flows has been
discovered. However, the verification by the theoretical side is
still absent since in many previous theoretical investigations the
vanishing flow was studied with a cluster charge-Z {\it independent}
global quantity
--- the so-called directed transverse momentum $<p_x^{dir}>$.
In addition, the isospin effect should be visible in (the time
evolution and the excitation function of) collective flows at low
SIS energies. The knowledge of the effect of isospin asymmetry on
both the mean field and the NNECS might be renewed under the new
version of the UrQMD. These studies are currently underway and will
be addressed in a forthcoming paper.

\section*{Acknowledgements}
We acknowledge support by the computing server C3S2 in Huzhou Teachers
College. The work is supported in part by the key project of the
Ministry of Education of China (No. 209053), the National Natural
Science Foundation of China (Nos. 10905021,10979023, 10979024), the
Zhejiang Provincial Natural Science Foundation of China (No.
Y6090210), and the Qian-Jiang Talents Project of Zhejiang Province
(No. 2010R10102).


\end{document}